\documentclass[12pt]{iopart}
\usepackage{graphicx}
\usepackage{url}
\usepackage{lineno}
\begin{document}

\title[Effect of the hadronic interaction models on the estimation of the CTA sensitivity]{Effect of the uncertainty in the hadronic interaction models on the estimation of the sensitivity of the Cherenkov Telescope Array}

%%%%%%%%%%%%%%%%%%%%%
%%% author list , IOP publishing format

\author{Michiko~Ohishi$^1$, Luan~Arbeletche$^2$, Vitor~de~Souza$^2$, Gernot~Maier$^3$, Konrad~Bernl\"ohr$^4$, Abelardo~Moralejo~Olaizola$^5$, Johan~Bregeon$^6$, Luisa~Arrabito$^6$, Takanori~Yoshikoshi$^1$}

% Addresses
\address{$^1$ Institute for Cosmic Ray Research, The University of Tokyo, Kashiwanoha 5-1-5 Kashiwa, Japan, 277-8582}
\ead{ohishi@icrr.u-tokyo.ac.jp}
\address{$^2$ Instituto de F\'{i}sica de S\~{a}o Carlos, Universidade de S\~{a}o Paulo, Av. Trabalhador S\~{a}o-carlense, 400 - CEP 13566-590, S\~{a}o Carlos, SP, Brazil}
\address{$^3$ Deutsches Elektronen-Synchrotron, Platanenallee 6, Zeuthen, 15738, Germany}
\address{$^4$ Max-Planck-Institut f\"{u}r Kernphysik, Saupfercheckweg 1, Heidelberg, 69117, Germany}
\address{$^5$ Institut de Fisica d'Altes Energies (IFAE), The Barcelona Institute of Science and Technology, Campus UAB, Bellaterra (Barcelona), 08193, Spain}
\address{$^6$ Laboratoire Univers et Particules de Montpellier, Universit\'{e} de Montpellier, CNRS/IN2P3, CC 72, Place Eug\'{e}ne Bataillon, Montpellier Cedex 5, F-34095, France}

%%%%%
%\address{IOP Publishing, Temple Circus, Temple Way, Bristol BS1 6HG, UK}
%\ead{submissions@iop.org}

\vspace{10pt}

\begin{abstract}
   Imaging Atmospheric Cherenkov Telescopes (IACTs) are ground-based indirect detectors for cosmic gamma rays with energies above tens of GeV. The major backgrounds for gamma-ray observations in IACTs are cosmic-ray charged particles. The capability to reject these backgrounds is the most important factor determining the gamma-ray sensitivity of IACT systems. Monte Carlo simulations are used to estimate the residual background rates and sensitivity of the systems during the design and construction phase. Uncertainties in the modeling of high-energy hadronic interactions of cosmic rays with nuclei in the air propagate into the estimates of residual background rates and subsequently into the estimated instrument sensitivity. We investigate the influence of the difference in the current hadronic interaction models on the estimated gamma-ray sensitivity of the Cherenkov Telescope Array using four interaction models (QGSJET-II-03, QGSJET-II-04, EPOS-LHC, and SIBYLL2.3c) implemented in the air shower simulation tool CORSIKA. Variations in background rates of up to a factor 2 with respect to QGSJET-II-03 are observed between the models, mainly due to differences in the $\pi^0$ production spectrum. These lead to $\sim$30\% differences in the estimated gamma-ray sensitivity in the 1~-~30~TeV region, assuming a 50-hour observation of a gamma-ray point-like source. The presented results also show that IACTs have a significant capability in the verification of hadronic interaction models.
\end{abstract}

\vspace{2pc}

\submitto{\jpg}
%\maketitle

%%%%%%%%%%%%%%%%%%%%%%%%%%%%%%%%%%%
\section{Introduction}
    
% GM's modifications were accepted and markups were removed by pyMergeChanges.py

Indirect cosmic ray detectors play an essential role in the observations of very-high-energy (VHE, $10^{11} - 10^{14}$ eV) and ultra-high-energy ($>10^{14}$ eV) cosmic gamma rays. Their large collection area ($>10^4$ m$^2$) is achieved by utilizing the extensive air shower (EAS) phenomena induced by the primary cosmic ray hitting the atmosphere. Since indirect detectors acquire the information of the secondary particles from EASs to learn about the primary cosmic ray, the accuracy of particle type identification is in principle limited compared with direct gamma-ray detectors on satellites and balloons. Therefore the precise understanding of the interaction between high-energy cosmic rays and nuclei in the air has been an important topic in the indirect cosmic ray experiments. It is essential for improving the accuracy of the measurement of the primary cosmic rays.

 Imaging Atmospheric Cherenkov Telescope (IACT) systems belong to this class of indirect cosmic ray detectors \footnote{IACTs can also detect Cherenkov photons from charged primary particles \cite{KiedaDC} and they partly work as direct cosmic ray detectors in the measurements of heavy nuclei \cite{HESSiron, VERITASiron}}. Thanks to their excellent particle identification capability, brought by the imaging method~\cite{ImagingHillas}, they are the most sensitive detectors of cosmic gamma rays in the VHE region. However, gamma rays account for a very small fraction ($<$ 1\% even for bright gamma-ray sources) of the triggered events in these systems. Even after selecting gamma-like events in the analysis, a large fraction of charged cosmic rays misidentified as gamma rays remains. Cosmic ray electrons produce electromagnetic (EM) showers which are very similar to those from gamma rays. Those cannot be distinguished on an event-by-event basis. Hadronic showers induced by cosmic ray protons and heavier nuclei also include EM showers as sub-structures, originated primarily from neutral pion ($\pi^0$) decay. Those sub-EM showers can mimic showers from gamma rays in IACT observations \cite{MAIER2007, SITAREK2018, Sobczy_ska_2007, Sobczy_ska_2015}. For these reasons, IACT systems do not achieve background-free gamma-ray observations and the amount of residual cosmic ray events is the most important factor determining the gamma-ray sensitivity of an IACT system.

In the derivation of the instrument sensitivity of current IACT systems, cosmic ray data (so-called {\it OFF-source} data) are used to estimate the residual backgrounds. Multivariate analysis (MVA) and machine learning (ML) techniques are commonly used in the analyses of IACTs for gamma/hadron separation, and they require background and signal data samples for the training. Cosmic ray data are frequently used as background events in this process, while Monte Carlo (MC) simulated gamma rays are used as signal events (see e.g., Refs.~\cite{MAGICRF, VERITASbdt2}). For this reason, cosmic ray protons and heavier nuclei are seldom simulated in the standard analysis of gamma-ray sources in current IACT systems. At the same time, assuring a good agreement between MC and data is essential for the gamma-ray observations by IACTs. Some of the MC input parameters (e.g., atmospheric transmission, night-sky-background, and optical throughput of the telescopes) are variable on a run-by-run basis. Checking the degree of matching between data and cosmic ray event simulations can still be useful for the constant monitoring of detector performance. Besides, recent machine learning techniques used for the efficient gamma/hadron separation perform only if the MC simulated events including background hadrons reproduce the real events accurately.  Therefore, efforts to achieve accurate simulations of hadronic components are still meaningful, even when the system is already in operation.

For IACT arrays in the design or construction phase, MC simulations of cosmic ray protons are essential to estimate the residual background level and gamma-ray sensitivity. There are two main sources of uncertainties affecting these hadron simulations: the assumed primary cosmic ray spectrum and the uncertainties in the modeling of hadronic interactions occurring in the air showers. Measurements in the VHE region by direct cosmic ray detectors such as AMS-02~\cite{AMS02-proton,AMS02-ep}, CALET~\cite{CALETproton,CALETe2}, DAMPE~\cite{DAMPEproton,DAMPEe} greatly reduced the uncertainties in the cosmic-ray spectrum. The difference of the cosmic-ray spectra measured by these detectors is within 10\% in the sub-TeV region. The impact on the estimation of the residual backgrounds is therefore limited. As a consequence, the uncertainties of the modeling of hadronic interactions become more important.
% In the following paragraph suggestions from Luan and Vitor was adopted with a little modifications

 The bulk of hadronic interactions cannot be described in terms of first principles in quantum field theory. Thus, phenomenological models have to be employed to describe cross sections and particle distributions in the final state~\cite{HighEnergyParticleDiffraction}. The free parameters of such models are compared and constrained to describe collider data. In particular, the Large Hadron Collider (LHC)~\cite{LHCMachine} has provided vital data that gave birth to the series of post-LHC models used in air shower simulations: QGSJET-II-04~\cite{QGSJET01}, EPOS LHC~\cite{EPOS-LHC}, and SIBYLL2.3~\cite{SIBYLL2.3}. Albeit these models share some commonalities, each implements a particular description of hadronic processes~\cite{AnnualReviewNucl}. This leads to a series of known discrepancies in the description of the microscopic reactions in a vast energy range~\cite{PhysRevD.98.083003}.
%UHECR description may not be needed
%iv) In the context of extensive air showers, the Pierre Auger and the Telescope Array collaborations have reported that none of the available models can describe the muon content in air showers (references here - those already in text). The large uncertainty related to the description of hadron-air interactions indeed misguides the interpretation of EAS data in terms of the primary mass compositions at the highest energies (perhaps cite 10.1103/PhysRevD.96.122003 and maybe references 4-7 therein ?).
Even at lower energies, those relevant to IACT measurements and below the LHC regime, studies have shown that some models can explain spectra of some types of secondary particles, but there is still no model that can consistently reproduce the measured spectra of all types of secondary particles~\cite{RibeiroPrado:2017os, Unger:2019nus}.
%studies have shown that none of these models can describe the measured secondary particle spectra~\cite{RibeiroPrado:2017os, Unger:2019nus}. As a consequence, the uncertainties related to the description of hadronic interactions are propagated to the evaluation of the Cherenkov light production in air showers in the IACT energy range~\cite{PARSONS2011, Parsons2019}. 

 %Simulation codes and hadronic interaction models have been updated over years and the performance of the IACT systems has also evolved in stages, the methodology for the validation of hadronic interaction models using the latest IACT system still needs to be studied.
  IACTs could potentially test and validate the interaction models as indirect cosmic ray detectors. Differences in the interaction models are expected to appear in various observables of IACTs, such as lateral distributions of Cherenkov photons on the ground, muon fluxes~\cite{PARSONS2011, Parsons2019, MITCHELL2019}, and shower image parameters. However, simulations of cosmic ray nuclei are performed in very limited cases (i.e., for the measurement of the cosmic ray electron, proton, and iron spectra~\cite{HessElectron2008, VERITASElectron2018,HEGRAproton, VERITAS_Iron,HESS_Iron}). Studies on the effect of the hadronic interaction modeling uncertainty using MC simulations with a realistic IACT detector response are limited. HEGRA performed MC simulations of proton and helium with two different interaction models~\cite{HEGRAproton}, modified RSM (radial-scaling model~\cite{RadialScalingModel}) in the ALTAI code~\cite{ALTAIcode} and HDPM (hadronic interactions inspired by the dual parton model~\cite{HDPM}) in CORSIKA~\cite{CORSIKA}. A good agreement between the two models was observed in the effective area and standard Hillas parameters \cite{ImagingHillas} distributions. H.E.S.S. performed proton simulations for the estimation of the background in the measurement of the cosmic ray electron spectrum~\cite{HessElectron2008}. Two interaction models were used, SIBYLL~\cite{SibyllHessEl} and QGSJET-II~\cite{QGSJETIIHessEl}. It was shown that the SIBYLL model produces more electron-like events than QGSJET-II and reaches better agreement with the data in the 1~-~4~TeV region. Proton simulations with an array layout similar to VERITAS were performed with two interaction models (QGSJET-01c~\cite{QGSJET01} and SIBYLL2.1~\cite{SIBYLL2.1Maier}) for the study of the nature of gamma-like proton events~\cite{MAIER2007}. A difference of up to 25\% in the collection area of gamma-like events was found.  While the simulation codes and hadronic interaction models have been updated over the last years, as well as the performance of the IACT systems, a methodology to validate these models using current and future observatories still needs to be developed.
   
 In gamma-ray observations, which is the primary scientific goal of IACTs, cosmic ray events are misidentified as signal events when they appear as a single electromagnetic shower. The nature of these gamma-like hadron events has been studied in previous works~\cite{MAIER2007, SITAREK2018, Sobczy_ska_2007, Sobczy_ska_2015}, and a sub-electromagnetic shower originating from a high-energy $\pi^0$ ($E_{\pi^0}$ close to $E_{\rm primary}$) is known to be a major source of these backgrounds. Therefore, predictions of the $\pi^0$ production spectrum, especially close to the primary energy, are expected to determine the rate of gamma-like events and affect the estimated gamma-ray sensitivity of IACT systems. 
 
 The Cherenkov Telescope Array (CTA) is a next-generation IACT project, where the plan is to construct two km-scale telescope arrays with 19 telescopes on the Northern site, and 99 telescopes on the Southern site \footnote{The total number of telescopes to be built on each site is subject to ongoing optimisations and we assume here the design configuration as discussed in \cite{ACHARYYA201935, MaierICRC2019}}. Given the large operating energy range (20~GeV to 300~TeV) of CTA, the array will be composed of three IACT classes, designated as the Large-, Medium- and Small-Sized Telescopes (LSTs, MSTs, and SSTs). The large-scale CTA arrays with wide field-of-view imaging cameras will allow a better coverage of the Cherenkov photons from EASs with respect to the current IACT arrays. They will provide more detailed views of air shower evolution, especially for hadronic showers where Cherenkov photons are scattered more widely and less symmetrically than in gamma-ray showers. For this reason, CTA is expected to have an excellent capability to validate hadronic interaction models, utilizing a combination of various observables such as shower image parameters, EM sub-shower rates,  muon fluxes, etc. 
 
  In this paper, we investigate the effect of the uncertainty in current interaction models on the sensitivity estimation of CTA, by testing four models (QGSJET-II-03~\cite{QGSJETII1}, QGSJET-II-04~\cite{QGSJETII2}, SIBYLL2.3c~\cite{SIBYLL2.3}, EPOS-LHC~\cite{EPOS-LHC}) implemented in the air shower simulation tool CORSIKA~\cite{CORSIKA} versions 6.99 and 7.69. In Sec. 2, simulations and analysis methods used in this work are described. The results of the comparison between interaction models are shown in Sec. 3. The possibility of the verification of the interaction models with CTA is discussed in Sec. 4.

\section{Simulation and Analysis}
    % GM's modifications were accepted and markups were removed by pyMergeChanges.py

  We perform two types of MC simulations to examine the effects of the difference in the hadronic interaction models on the observables of CTA and the resulting gamma-ray sensitivity. The first type of simulation is performed in order to check the properties of the EAS secondary particles and does not include the detector response. The second type of MC simulation aims at investigating the effect on the gamma-ray sensitivity of CTA, and includes the detector response and array configuration. In both schemes, QGSJET-II-03 in CORSIKA 6.99 and post-LHC models in CORSIKA 7.69 (QGSJET-II-04, SIBYLL2.3c, EPOS-LHC) were used as high-energy ($E > 80$ GeV) interaction models in proton-induced showers.
  QGSJET-II-03 is the model used to derive the current public Instrument Response Functions (IRFs) of CTA~\cite{CTApublicIRFprod3bv2}.
  As for the low energy model for $E < 80$~GeV, a single fixed model, UrQMD1.3cr [46] was used. Therefore, the evaluation of the influence of the choice of hadronic interaction model in this study is limited to that of the high energy models.
  %The choice of low energy models (UrQMD1.3cr~\cite{UrQMD}, FLUKA~\cite{FLUKA1, FLUKA2}) used below 80~GeV is not expected to have a strong impact on the estimation of CTA sensitivity~\cite{Sobczy_ska_2015}, 
%  A single low-energy model, UrQMD1.3cr~\cite{UrQMD}, was used for all simulations in this work.

\subsection{Air shower simulation without CTA detector response}
 In the simulation without detector response, the environmental setting such as the atmosphere profile and the geomagnetic fields were set to be the same as in the simulation of the CTA South site in Chile~\cite{HASSAN201776}. Track information of particles (particle type, energy, coordinate of start- and end-point of the particle trajectory) in the simulated showers were extracted using the CORSIKA PLOTSH option~\cite{CORSIKAmanual}. The primary proton energy was set to be mono-energetic at energies of 0.1, 0.316, 1.0, 3.16, and 10~TeV and approximately $10^5$ events were simulated for each energy. The target nucleus was fixed as nitrogen in order to make the comparison of the models easier. The lower energy limit of tracked EM particles ($e^-$, $e^+$, $\gamma$, $\pi^0$) was set to be 0.1\% of the primary proton energy, $E_p$, to suppress large-size track outputs in high-energy showers. Hadrons and muons were tracked until their kinetic energies become below 300~MeV and 100~MeV, respectively. From this track information, $\pi^0$ particles whose energies are within one decade of the primary proton energy ($0.1E_p \leq E_{\pi^0}\leq 1.0E_p$) were collected in order to examine the energy distribution near $E_p$. The fraction of the primary particle energy carried by EM particles ($\gamma$, $e^-$, and $e^+$) was calculated after the third interaction. Here all the processes (hadronic, electromagnetic, or decay) with the appearance of a new particle with energy above 0.1\% of $E_p$ are counted as one step of the interaction. The products just after the first interaction provide useful information to test the models as well~\cite{Parsons2019}, but in many cases nucleons carry a significant fraction of energy at this early stage, and the nature of the shower is not well determined yet (in particular its similarity to a shower from a gamma ray). For this reason, we used the products at the third interaction where the shower has evolved sufficiently and the energy fraction carried by the EM components becomes almost saturated for TeV primaries.

\subsection{Simulations including CTA detector response}
 The simulations including detector response followed previous CTA MC simulation studies~\cite{sim_telarray}. The array configuration and detector response in this work were set to be that of the CTA MC Production {\it Prod3b}~\cite{ACHARYYA201935, MaierICRC2019}. The South site array configuration of 99 telescopes with three different telescope types was chosen as it covers the full energy range. In the derivation of the IRFs and the sensitivity curve of a gamma-ray detector, MC simulation datasets of gamma rays (signal), electrons, and protons (backgrounds) are required. Common gamma-ray and electron MC datasets were used in the tests of all hadronic interaction models. The proton (QGSJET-II-03), electron and gamma-ray MC datasets were produced on the European Grid Infrastructure (EGI) with CORSIKA 6.99. The rest of the proton MC datasets with post-LHC models were produced with identical {\it Prod3b} setup, using computing resources in Japan consisting of 2,700 $\times$ 2.2GHz CPU cores. Input parameters for the MC simulations used in the derivation of IRFs are shown in Tables~\ref{tab:SimParameter}~and~\ref{tab:MCSimDataset}. Spectra of background cosmic ray protons and electrons are set to be the same as those used in the derivation of the public IRFs \cite{CTApublicIRFprod3bv2}. They were obtained by fitting the results of the previous direct/indirect measurements (ATIC~\cite{ATIC2009}, Fermi-LAT~\cite{FermiLATelectron2010}, H.E.S.S.~\cite{HessElectron2008} \cite{HESSelectron2009}, MAGIC~\cite{MAGICelectron}) with the following parameterizations of the spectra:\\
Proton:
\begin{equation}
    \frac{dN}{dE}=C_p(E/{\rm TeV})^{-\Gamma_p},
    \label{CRspec_proton}
\end{equation}

\[ C_p=9.8\times10^{-6}\ {\rm cm}^{-2}\,{\rm s}^{-1}\,{\rm sr}^{-1}\,{\rm TeV}^{-1},\,\Gamma_p=2.62. \]
 Electron:
\begin{equation} 
    \frac{dN}{dE} = C_e(E/{\rm TeV})^{-\Gamma_e} \times\left[ 1+f\times\left(\exp(G(E))-1\right)\right],
    \label{CRspec_electron}
\end{equation}

\[ G(E)=\exp\left(-\frac{\left(\log_{10}(E/{\rm TeV})-\mu\right)^2}{2\sigma^2}\right), \]

\[ C_e=2.385\times10^{-9}\ {\rm cm}^{-2}\,{\rm s}^{-1}{\rm sr}^{-1}{\rm TeV}^{-1}, \] 
\[ \Gamma_e=3.43,\,f=1.950,\,\mu=-0.101,\,\sigma=0.741. \]
where a single power-law spectrum for proton, and a power-law with an additional Gaussian hump for electron are assumed. $f, \mu, \sigma$ correspond to magnitude, center energy (in $\log_{10}E$), and standard deviation of the Gaussian, respectively.  
Simulation of nuclei heavier than proton is not required in the derivation of the IRFs, since it is known that heavy nuclei contribute little to the production of gamma-like background events~\cite{HessElectron2008, VERITASElectron2018, SITAREK2018, sim_telarray}. However, the effect of the contribution from helium is checked in Sec. 4. We used the single power-law energy spectrum shown in Eq.~(\ref{CRspec_proton}) with the following parameters for helium:
\[ C_{\rm He} =6.9\times10^{-6}\ {\rm cm}^{-2}{\rm s}^{-1}{\rm sr}^{-1}{\rm TeV}^{-1},\, \Gamma_{\rm He}=2.55, \] 
where $E$ corresponds to the energy per nucleus.
%In this work additional proton QGSJETII-03 MC data produced on the computer cluster in Japan was also used to improve the residual proton event statistics in the derivation of the sensitivity curve, after checking the consistency of the MC data consistency.
 The analysis method and software used in the derivation of the CTA IRFs (EventDisplay~\cite{EventDisplay}, v500-rc04) are similar to those used to produce the {\it Prod3b} ones. The basic flow of the analysis is as follows: 1)~camera image cleaning to remove night sky background\footnote{We used the optimized next-neighbour cleaning method \cite{OptNNcleaning}, where the cleaning threshold is determined by the fake image probability (set to be 0.05\% in this work) from the fluctuation of night-sky-background photons} and extraction of shower image characteristics (calculation of Hillas parameters);  2)~stereo reconstruction of shower parameters (energy, arrival direction, shower core position, etc.);  3) training of the Boosted Decision Trees (BDT)~\cite{Breiman1983ClassificationAR, Freund1999ASI, VERITASbdt2} for gamma/hadron classification using gamma and proton simulation datasets; 4) optimization of the BDT output cut value; 5) estimation of the gamma-ray sensitivity and other performance parameters using the signal events and the residual background events surviving the optimized cut. 
 All datasets of the four interaction models were treated in the same way, repeating the procedure mentioned above. The analysis was optimized for the observation of point-like sources.

\begin{table}
    \centering
    \begin{tabular}{cc}
    \hline
    Parameter & Value \\
    \hline
    Site & Paranal (Chile), 2150 m a.s.l. \\
    Array configuration & Prod3b \cite{MaierICRC2019}  \\
    Zenith angle & 20 deg \\
    Azimuthal direction & North + South  \\
    Spectral index &  -2.0 \\
    Viewcone & 10 deg \\
    Core range & 2000 m (point), 2500 m (diffuse) \\
    Number of shower reuse & 20 \\
    \hline
    \end{tabular}
    \caption{Some of the parameters used in the MC simulation with CTA detector response. In order to reduce the computing cost, re-use of showers is applied using CORSIKA CSCAT option \cite{CORSIKAmanual}. Superposition of 20 arrays is simulated with random position offsets on an event-by-event basis, and the outputs are counted as 20 independent showers with different core positions. Simulated arrival directions correspond to a circular cone with a half-opening angle of 10 degrees. Core range corresponds to the radius of a circle in which the array positions are randomly set.  Spectral index is fixed as -2.0 in the simulation, but re-weighted in the analysis to fit the functions shown in the text as Eq.~(\ref{CRspec_proton})~and~(\ref{CRspec_electron}).}
    \label{tab:SimParameter}
\end{table}

\begin{table}
    \centering
    \begin{tabular}{|l|l|l|}
    \hline
    gamma, 0.003 - 330 TeV& electron, 0.003 - 330 TeV  & proton, 0.004 - 600 TeV\\
    $1\times10^9$ events & $2 \times 10^8$ events & $1.2\times10^{11}$ events per model \\
    \hline
    & & C6.99 + QGSJET-II-03 \\ \cline{3-3}
    diffuse, point  &   & C7.69 + QGSJET-II-04 \\ \cline{3-3}
    C7.69 + QGSJET-II-03   & C6.99 + QGSJET-II-03 & C7.69 + EPOS-LHC \\ \cline{3-3}
    & & C7.69 + SIBYLL2.3c \\ \cline{3-3}
    \hline
    \end{tabular}
    \caption{Sizes and energy ranges of the MC simulation datasets used in the derivation of the IRFs. The gamma-ray and electron MC datasets are the same in all sensitivity calculations. C6.99/C7.69 denote the CORSIKA versions 6.99 and 7.69.}
    \label{tab:MCSimDataset}
\end{table}

\section{Results}
    % GM's modifications were accepted and markups were removed by pyMergeChanges.py 
\subsection{Air shower simulation without detector response}
\subsubsection{Energy spectrum of neutral pions in the shower}
\begin{figure}[t]
\centering
     \includegraphics[width=0.6\linewidth, keepaspectratio]{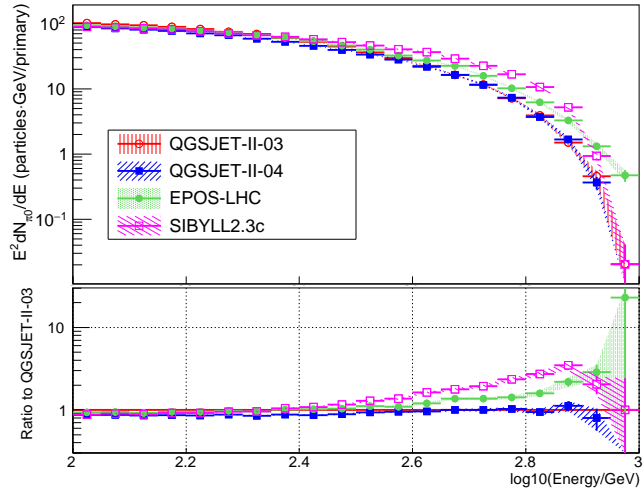}
     \caption{ (Top) energy spectra of $\pi^0$s in 1~TeV primary proton showers for the four interaction models using CORSIKA 6.99/7.69. Hatched areas correspond to statistical error bands, common to all figures in this work. All neutral pions in the showers are counted. (Bottom) ratio of spectra relative to the QGSJET-II-03 spectrum. }
    \label{fig:pi0spectrum}
\end{figure}

 The high-energy part of the spectrum of the secondary $\pi^0$ close to the primary proton energy is expected to affect the probability of the production of gamma-like background events. Hard spectra close to the primary proton energy lead to more frequent production of gamma-like events and vice versa. Figure~\ref{fig:pi0spectrum} shows the $\pi^0$ energy spectrum obtained with the four interaction models for 1~TeV mono-energetic primary protons. One decade of energy below the primary energy $E_p$ is shown. All $\pi^0$s produced in the shower are counted.
The differences between the models are at the 13\% level at $0.1E_p$ (with respect to QGSJET-II-03) and become larger at higher energies (up to 180\% at 0.8$E_p$). The two QGSJET-II models produce similar spectra above 0.4$E_p$, and both are relatively soft. SIBYLL2.3c produces the hardest spectrum up to 0.65$E_p$, which then steeply cuts off near the primary proton energy. EPOS-LHC produces the hardest spectrum. It is expected from these features that EPOS-LHC and SIBYLL2.3c will produce more gamma-like background events than the two QGSJET-II models at energies around 1~TeV.

\subsubsection{Energy fraction of the electromagnetic components}
\begin{figure}[htbp]
    \centering
    \includegraphics[width=0.6\linewidth, keepaspectratio]{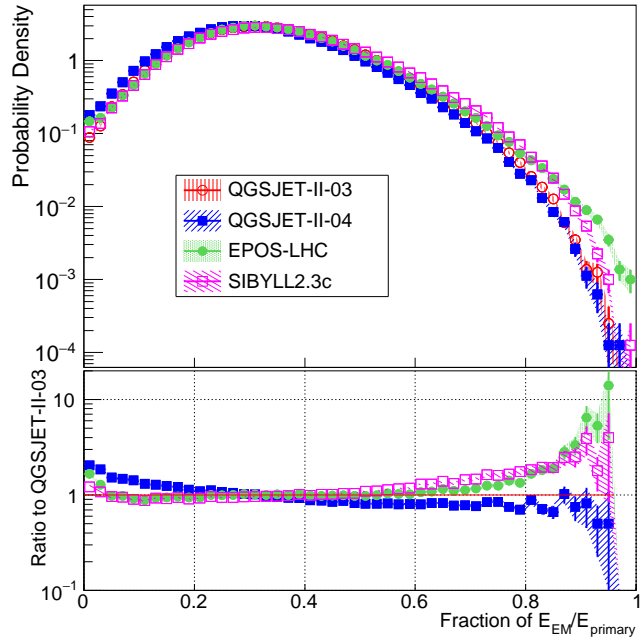}
    \caption{ (Top) probability density distribution of the energy fraction in the electromagnetic component ($\gamma$, $e^-$, $e^+$) for the four interaction models for primary proton energies of 1~TeV. (Bottom) ratio to QGSJET-II-03.}  
    \label{fig:EMfraction}
\end{figure}

 The gamma-like nature of an air shower can be interpreted as the similarity of a background event to an electromagnetic shower induced by a gamma ray. The energy fraction found in the electromagnetic component ($\gamma$, $e^-$, $e^+$) of the shower is expected to correlate with its gamma-like nature. For gamma-ray primaries, this fraction is close to 100\%. Figure~\ref{fig:EMfraction} shows the probability distribution of the energy fraction in the electromagnetic component after the third interaction for 1~TeV mono-energetic primary protons in the four models. All models produce a peak around $1/3$ with broad tails caused by event-by-event differences of the secondary products. It should be noted that this fraction depends on the primary proton energy and increases towards high energy. The probability of interaction between secondary $\pi^{\pm}$s and nuclei in the air is determined by their lifetimes, which in turn depend on their energies. In the low energy limit, $\pi^{\pm}$s decay into muons\footnote{Kaons ($K^{\pm}$) also contribute to the muon production, but their contribution is an order of magnitude lower than that of $\pi^\pm$ in the energy budget} before interacting with the air, and the energy fraction of the electromagnetic component from $\pi^0$ becomes closer to $1/3$. There are additional EM components from muon decays which increase towards the lowest energy, but in the energy budget they reach up to 1/3 of $\pi^0$ component even if all the muons decay before reaching the ground. 
 
\begin{figure}[htbp]
    \center
    \includegraphics[width=0.9\linewidth,keepaspectratio]{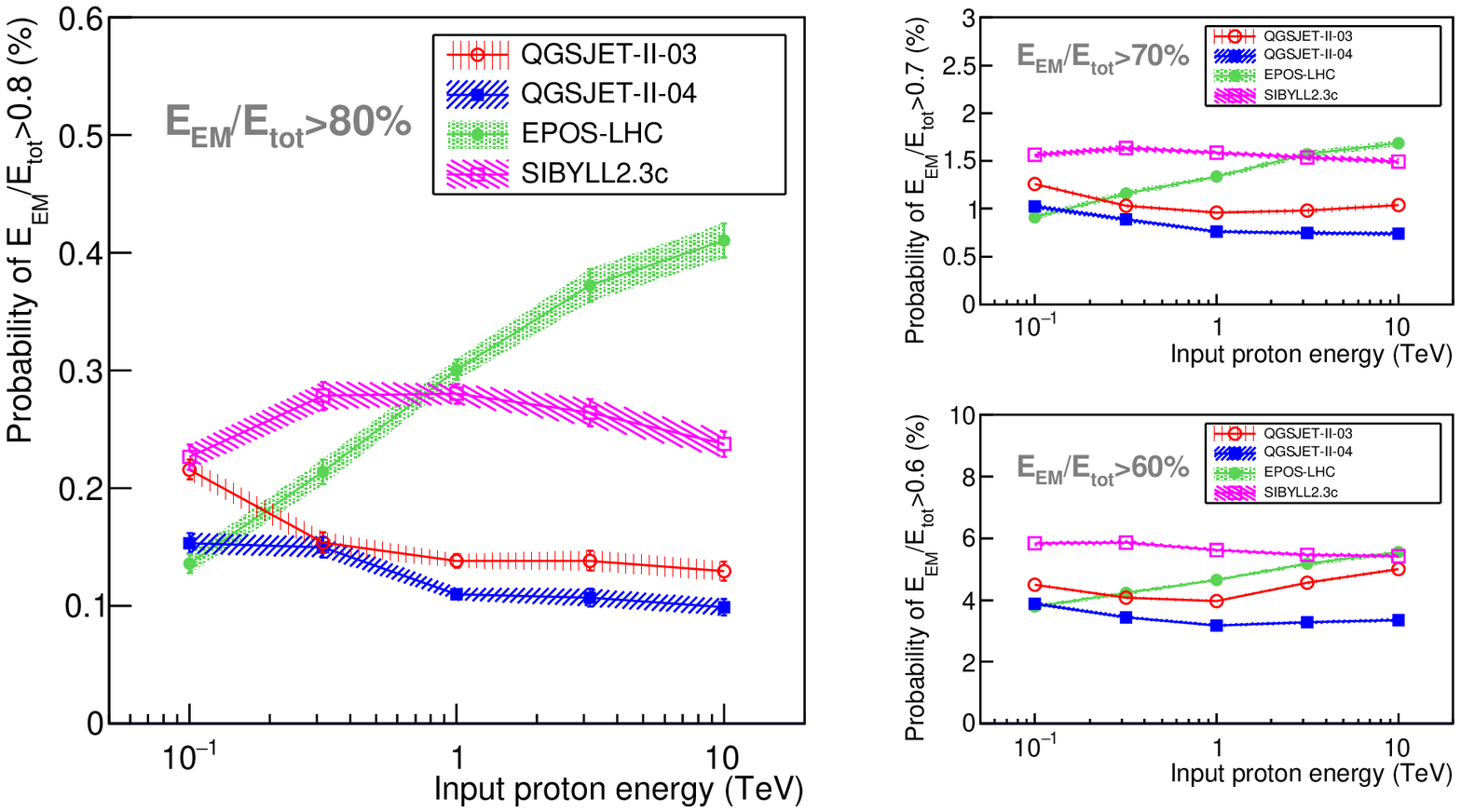}
    \caption{Energy dependence of the probability of high electromagnetic fraction events for the four interaction models. Cases with different thresholds in EM fraction are shown: (Left) $80\%$, (right top) $70$\%, (right bottom) $60$\%.  Relation between models varies depending on the threshold value. }  
    \label{fig:EMfraction_Edep}
\end{figure}

 The probability of high EM-fraction events is a good indicator for the production rate of gamma-like events. As expected from the $\pi^0$ spectrum, EPOS-LHC and SIBYLL2.3c indicate higher EM-fraction probabilities than the two QGSJET-II models. Figure~\ref{fig:EMfraction_Edep} shows the energy dependence of the probability of high electromagnetic fraction (EM) events ($E_{\rm EM}/E_{\rm primary} > 80\%$, $70\%$ and $60\%$) for primary proton energies from 0.1 to 10~TeV. An energy dependence is observed for this fraction for EPOS-LHC, decreasing towards lower energies and crossing with SIBYLL2.3c around 1~TeV in the case of $E_{\rm EM}/E_{\rm primary} > 80\%$. The energy where the SIBYLL2.3c and EPOS-LHC cross depends on the selection of the threshold in energy fraction. Lowering this threshold value makes the crossing point of EPOS-LHC and SIBYLL2.3c higher, and the difference between QGSJET-II-03 and QGSJET-II-04 larger. The crossing point for the case of $E_{\rm EM}/E_{\rm primary}>80\%$ is close to the results of the differential sensitivity study presented in the next section. It should also be noted that smaller IACT arrays tend to have higher probabilities of misidentifying protons as gamma rays even if the energy fraction in electromagnetic components is low, as they might observe an electromagnetic sub-shower only and not the entire air shower. Therefore a fixed threshold value for this EM fraction over a broad energy range may be over-simplified considering the design of CTA.
 
    %% GM's modifications were accepted and markups were removed by pyMergeChanges.py

\subsection{Simulations including the CTA detector response}
 The simulation including the CTA detector response aims at investigating how the differences in interaction models affect the observables of IACTs and their sensitivity to gamma-ray sources. In this section, we discuss energy scale\footnote{The energy scale here is the relation between the simulated proton energy and the reconstructed energy obtained from the Cherenkov image analysis}, proton shower rates, collection area, distributions of basic shower parameters, Multivariate Analysis (MVA) parameters for gamma/hadron separation, and gamma-ray sensitivity. A point-like gamma-ray source is assumed in the derivation of the gamma-ray sensitivity.

 \subsubsection{Relation between reconstructed and simulated energy and the impact on proton shower rate} 
 
\begin{figure}[ht]
    \centering
    \includegraphics[width=0.8\linewidth,keepaspectratio]{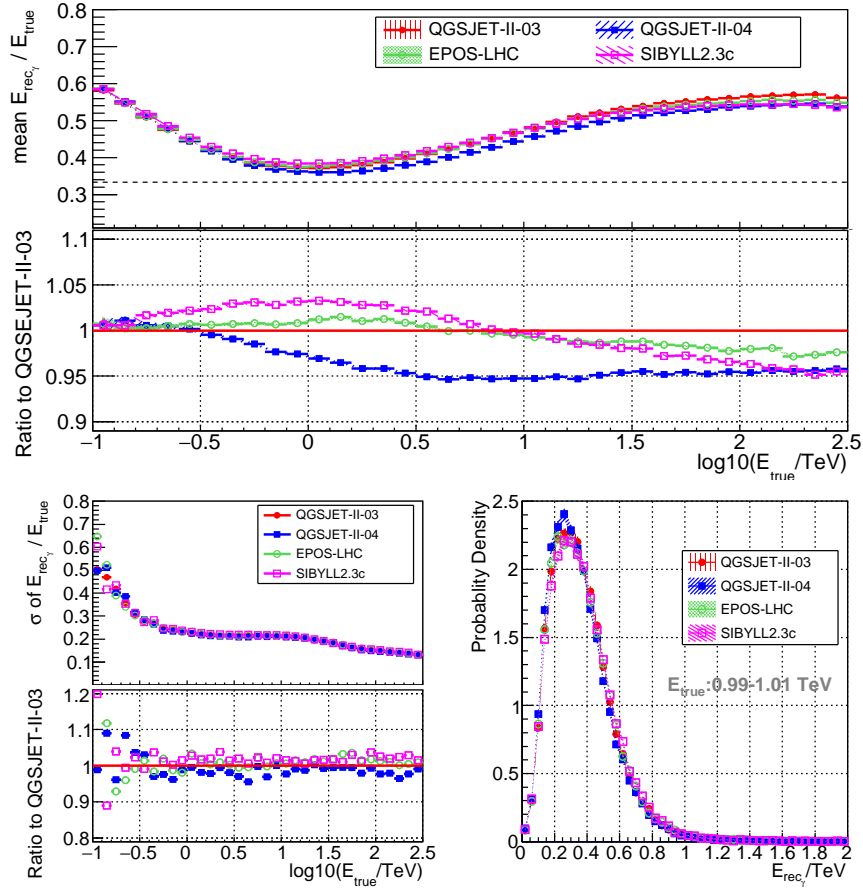}
    \caption{ (Top) ratio of the mean reconstructed energy ($E_{{\rm rec}_{\gamma}}$) to the true simulated energy ($E_{\rm true}$) for the four interaction models, after image cleaning and before the selection of gamma-like events. Energy is reconstructed assuming the incoming particles are gamma rays, which leads to the systematically lower $E_{{\rm rec}_{\gamma}}$ compared to $E_{\rm true}$. Events with telescope multiplicity $\geq$ 4 are plotted, regardless of the type of the telescope. The horizontal dashed line in the upper panel corresponds to 1/3. Ratio of mean $E_{{\rm rec}_{\gamma}}$  in each energy bin with respect to the QGSJET-II-03 model is also shown in the lower panel. (Bottom left) standard deviation (SD) of the distributions of $E_{{\rm rec}_{\gamma}}$ divided by ${E_{\rm true}}$, and the corresponding ratio to the QGSJET-II-03 model. (Bottom right) $E_{{\rm rec}_{\gamma}}$ distributions for the protons with $E_{\rm true}$ of 0.99-1.01~TeV (a substitute of mono-energetic proton of 1~TeV, events are re-weighted to make an uniform energy distribution within this narrow energy band). }
    \label{fig:ReconstructedE}
\end{figure}

 Since the major contributors for Cherenkov photon emission in extensive air showers are electrons and positrons, the energy fraction carried by the electromagnetic component affects the energy estimation in IACT analyses. The difference in hadronic interaction models can be seen in the relation between the average reconstructed gamma-ray energy ($E_{{\rm rec}_{\gamma}}$; proton events are reconstructed assuming that they are gamma-ray events) and the true simulated proton energy ($E_{\rm true}$). This difference subsequently modifies the rate of proton showers surviving selection cuts as a function of $E_{{\rm rec}_\gamma}$. Figure~\ref{fig:ReconstructedE} shows the relation between $E_{\rm rec_{\gamma}}$ and $E_{\rm true}$ for proton events produced with the four interaction models, after image cleaning, before the selection of gamma-like events and after requiring a telescope multiplicity $\geq$ 4 regardless of the type of the telescope. The average reconstructed energy is close to $E_{{\rm rec}_{\gamma}}=\frac{1}{3}E_{\rm true}$ around 1~TeV. (It should be noted that $E_{{\rm rec}_{\gamma}}/E_{\rm true}$ becomes larger after the selection of gamma-like events, since the gamma-like events have a higher energy fraction in the electromagnetic components.) It can be seen in Fig.~\ref{fig:ReconstructedE} that $E_{{\rm rec}_{\gamma}}$ is not proportional to $E_{\rm true}$ and increases towards high energy and reaches $\frac{1}{2}E_{\rm true}$ at 30~TeV. The increase in the low energy region ($<$ 300~GeV) is due to the effect of the bias in the selection of events with an upward fluctuation in the number of Cherenkov photons, and to the additional EM components from muon decay. In the 1~-~10~TeV region there is a 5-7\% difference seen in $E_{{\rm rec}_{\gamma}}$ between the models. QGSJET-II-04 is lower than others, reflecting that it has more events with a low electromagnetic fraction, as shown in Fig.~\ref{fig:EMfraction}. The 5-7\% difference in $E_{{\rm rec}_{\gamma}}$ propagates to a 8-12\% difference in the proton shower rate, assuming a primary cosmic ray proton spectrum index of $-2.62$. Differences in proton shower rate in turn affect sensitivity to gamma-ray sources estimates. Figure \ref{fig:ReconstructedE} also shows the standard deviation (SD) of $E_{{\rm rec}_{\gamma}}$ with respect to $E_{\rm true}$, along with $E_{{\rm rec}_{\gamma}}$ distribution for the protons with $E_{\rm true}$ of $0.99 - 1.01$~TeV (a substitute of mono-energetic proton of 1 TeV). Proton events have more complex images than gamma events and it makes  reconstruction of the shower core positions more difficult ( $\sim$150~m as a 68\% containment radius for 1~TeV proton). The limited accuracy of the estimated impact parameter is propagated to the energy estimation. Along with the effect of the variation in the products in the hadronic interaction, $E_{{\rm rec}_{\gamma}}$ has a large distribution width, 0.2 TeV in SD for $E_{\rm true}$ 1~TeV proton. The uncertainty in the reconstructed energy is estimated on an event-by-event basis and it is used in the MVA analysis for gamma/hadron separation.
 
\subsubsection{Collection area}
\begin{figure}[ht]  
    \centering
    \includegraphics[width=0.8\linewidth,keepaspectratio]{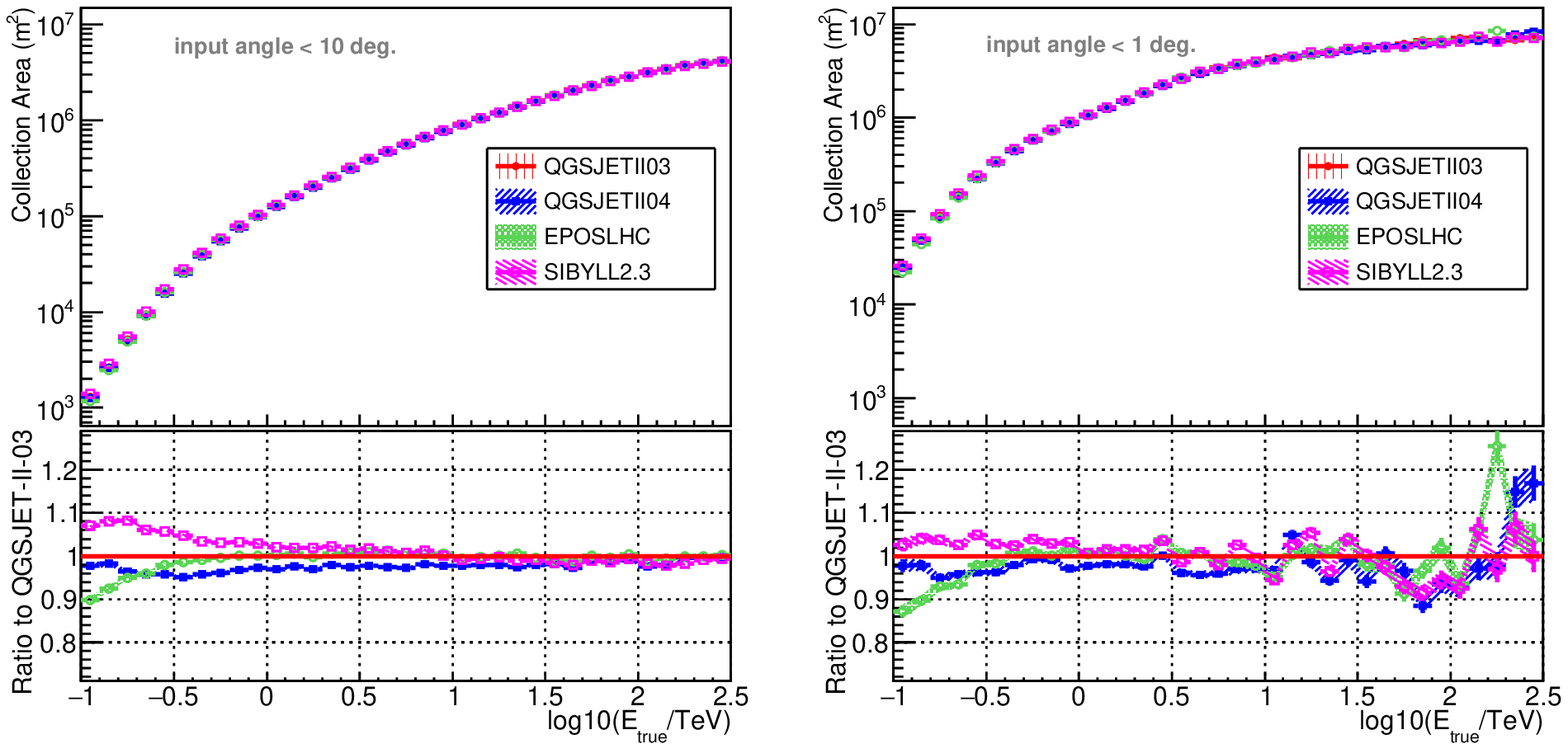}
    \caption{Collection areas with respect to $E_{\rm true}$ for the four interaction models, with the same event selection as the energy scale plot. Note that this collection area was calculated for the diffuse protons and the resulting value depends on the range of the input angle. (Left) input angle $<$ 10 degrees, (right) input angle $<$ 1 degree. The range is defined as a half-opening angle of a circular cone. }
    \label{fig:Aeff}
\end{figure}

Figure \ref{fig:Aeff} shows the proton collection area for the four interaction models with respect to $E_{\rm true}$. The same event selection as used in the energy scale plot was applied. Here the effective area is simply calculated as $S_{\rm scat}\times\frac{N_{\rm accepted}}{N_{\rm simulated}}$, where $S_{\rm scat}$ is the area in which the shower core positions were scattered in the simulation, and $N_{\rm simulated}$ and $N_{\rm accepted}$ are numbers of simulated events and survived events after event selection. Since the incoming protons have a uniform distribution in their directions, the calculated effective area strongly depends on the range of the input angle.  Figure \ref{fig:Aeff} shows two cases for the input angle range,  10 degrees (left) and 1 degree (right). The range is defined as half-opening angle of a circular cone.  

 As for the 10 degrees case, the difference among models is small in high energy regions (less than 5\% above 3~TeV) and becomes larger towards low-energy, reaching 10\% level at 0.3~TeV. This difference is expected to reflect the combined effect of difference in the production of EM components and spatial pattern of Cherenkov photons on the ground. In the 1 degree case,  the effect of different field-of-view sizes between the telescopes is eliminated by targeting the protons within a narrow cone around the telescope optical axis. The overall value of the collection area becomes higher, and the rapid increase towards higher energies is mitigated compared with the 10 degrees case. As for the relations between the models, it shows a similar trend to the 10 degrees case (with just a few percent drop for SIBYLL2.3c and EPOS-LHC at the lowest energy).  This feature suggests that the difference of the angular distributions of secondary particles between the models is not so large to affect the collection area significantly.

\subsubsection{Distributions of basic shower parameters} 
\begin{figure}[ht]
    \centering
    \includegraphics[width=0.6\linewidth,keepaspectratio]{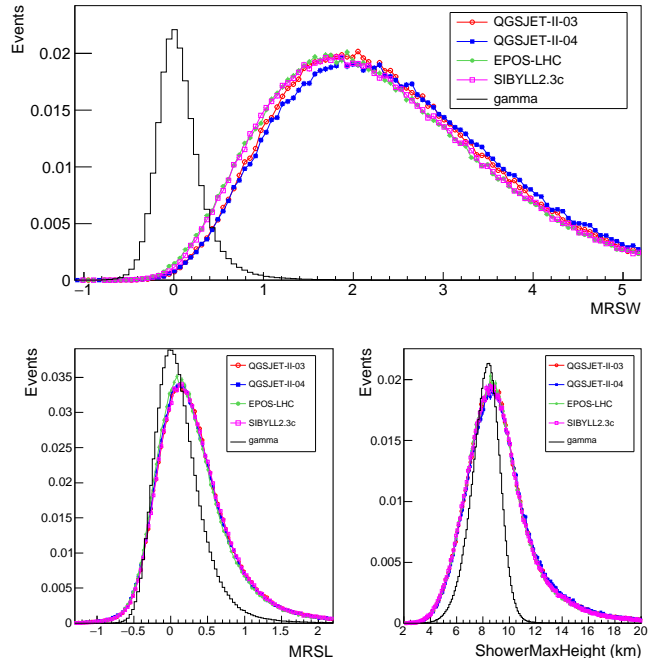}
    \caption{Distributions of basic shower parameters for proton events produced with the four interaction models, for energies $E_{{\rm rec}_{\gamma}}$ of 1~-~10~TeV. Histograms are normalized by areas (number of the accepted events). The difference in shower rates from the effect shown in Fig.~\ref{fig:ReconstructedE} is not included. (Top) {\it MRSW (Mean Reduced Scaled Width)}, (bottom left) {\it MRSL (Mean Reduced Scaled Length)}, (bottom right) Height of shower maximum, measured from the observation level.  } %Proton events were re-weighted so that simulated events reproduce the spectrum shown in Eq.~\ref{CRspec_proton}.
%Differences in shower rate of $\sim$10$\%$ level is included in this figure. Gamma-ray histograms are in arbitrary scale. }
    \label{fig:BasicShowerParameters}
\end{figure}

Properties of secondary particles (angular distributions, energy spectra, particle types) produced in shower interactions determine the topology of shower images observed in IACTs. The typical parameters representing the shower features are {\it Width}, {\it Length}, and the height of shower maximum (here called {\it Shower maximum}). {\it Width} and {\it Length} correspond respectively to the observed lateral and longitudinal size of a shower image in a single telescope camera. Combining the parameters from each telescope and taking into account the expected dependency on the impact parameter and intensity (sum of photo-electrons of a shower image), the {\it Mean Reduced Scaled  Width (MRSW)} and {\it Mean Reduced Scaled Length (MRSL)} \cite{HessCrab} are calculated,
\begin{equation}
  MRSW = \frac{1}{N_{\rm tel}}\sum_{i}^{N_{\rm tel}}\frac{Width_i - Width_{\rm expected}( R_i, I_i)}{\sigma_{\rm expected}( R_i,I_i)},
\end{equation}
where $N_{\rm tel}$ is the number of triggered telescopes, and $R_i$ and $I_i$ are the impact parameter and intensity of the $i$-th image. The expectation value is extracted from look-up tables prepared from MC gamma-ray events in the relevant zenith and azimuth angle region. The {\it MRSL} is calculated similarly.
Figure~\ref{fig:BasicShowerParameters} shows the distributions of the shower parameters, {\it MRSW}, {\it MRSL}, and the {\it Shower maximum} for the four interaction models in the $1 < E_{{\rm rec}_{\gamma}} < 10$~TeV band.
A telescope multiplicity of 4 for any type of telescopes ($N_{\rm LST}=4$ or $N_{\rm MST}\geq 4$ or $ N_{\rm SST}\geq4$) is    required (tighter event selection than in the case of the energy scale; this multiplicity criterion is used in the following analysis steps leading to the sensitivity curve derivation). An offset angle\footnote{The offset angle is defined as the angular distance between the reconstructed arrival position and the camera center} of less than 1.5 degrees is also required in order to select well-contained events in the imaging camera. The distributions are normalized by their areas so that the differences in shower rate between models are not considered.  
{\it MRSW} is the most effective parameter to distinguish between gamma and hadron primaries at TeV energies.
In the {\it MRSW}$<$1.0 region (or gamma-like region), EPOS-LHC and SIBYLL2.3c produce more events than the two QGSJET-II models, as expected. The QGSJET-II-03 and 04 distributions in this region are similar. Excluding the difference in the shower rate, the ratios of events to QGSJET-II-03 in the {\it MRSW}$<$1.0 region are: $-9\pm1$\% (QGSJET-II-04), $31\pm1$\% (EPOS-LHC) and 29$\pm$1\% (SIBYLL2.3c). In the {\it MRSW}$>$1.0 region (proton-like region), a difference can be seen between QGSJET-II-03 and 04, where the peak position of the QGSJET-II-04 distribution deviates towards higher {\it MRSW} values. For the {\it MRSL} and the {\it Shower maximum} parameters a good agreement between the models is seen. 

\subsubsection{Distributions of MVA parameter}
\begin{figure}[ht]
    \centering
    \includegraphics[width=0.8\linewidth, keepaspectratio]{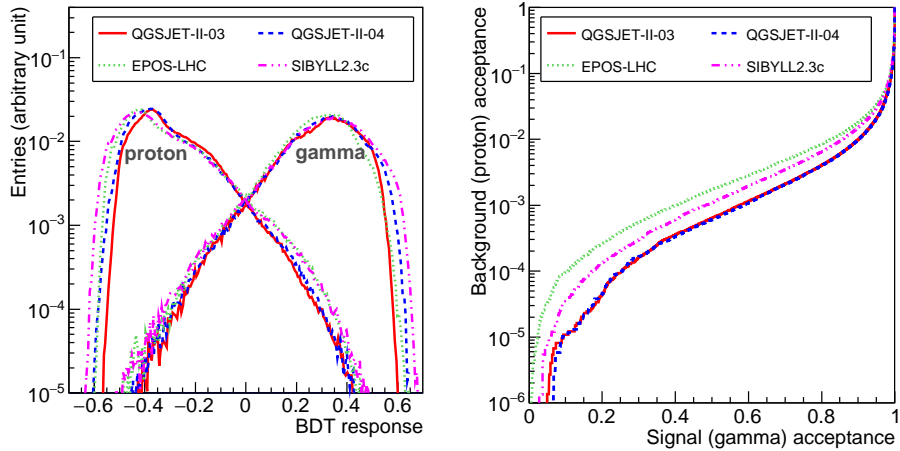}
    \caption{(Left) BDT distributions of gamma (signal) and proton (background) for the four interaction models. BDTs are trained with each interaction model. Data subset of $0.0 \leq \log_{10}(E/{\rm TeV}) \leq 0.75$ and offset angle $<0.5$ degrees is shown. (Right) relation between the acceptance of  background and signal, obtained by scans of cut value on the BDT response shown in the left figure. Comparison of the background acceptance at a certain signal acceptance gives a measure of the degree of separation between gamma and proton. }
   \label{fig:BDT_Sec3_gammaAna}
\end{figure}

The basic shower parameters are used in a multivariate analysis (MVA) to produce a single indicator ({\it gammaness}) for gamma-hadron classification. The Boosted Decision Trees (BDT) technique implemented in ROOT TMVA~\cite{hoecker2007tmva} was used with 11 input parameters, including those shown in Fig.~\ref{fig:BasicShowerParameters}.
% 11 parameters : MSCW, MSCL, log10(EChi2S), EmissionHeight, log10(EmissionHeightChi2), log10(SizeSecondMax), NImages_Ttype[3], dES, log10(DispDiff)
 MC datasets of diffuse gamma rays and protons are used as signal and background samples for the training of BDTs. MC datasets are divided into 54 subsets according to their $E_{{\rm rec}_{\gamma}}$ (9 bins) and offset angles (6 bins), and training was performed with each subset. In evaluating the BDT response, a relevant BDT is selected from the 54 subsets, based on the energy and offset angle of the event.   
 Figure~\ref{fig:BDT_Sec3_gammaAna} shows the BDT distributions of gamma (signal) and proton (background) events produced with the four interaction models in one of the subsets (1.0~TeV $ \leq E_{{\rm rec}_{\gamma}} \leq 5.6$~TeV and offset angle $<$~0.5 degrees). Since the BDTs are trained with each interaction model, the distributions of BDT response for gamma rays are also different from model to model. The proton BDT distributions for EPOS-LHC and SIBYLL2.3c have more tail components in the gamma-like (BDT~$>0$) region. However, a comparison of the models using proton BDT distributions should take into account the signal acceptance. We obtained a fraction of events which survives an event selection with a cut value ($\zeta_{\rm thres}$) on BDT response ($\zeta$), as a function of $\zeta_{\rm thres}$: $C(\zeta_{\rm thres})=N(\zeta >\zeta_{\rm thres})/N_{\rm total}$, where $N$ is the number of events. This function was obtained both for proton (background) and gamma (signal). The right panel of Fig.~\ref{fig:BDT_Sec3_gammaAna} shows the relation between $C_{\rm background}(\zeta_{\rm thres})$ and $C_{\rm signal}(\zeta_{\rm thres})$, obtained with scans on the value of $\zeta_{\rm thres}$. Comparison of the background acceptances at a certain signal acceptance gives a measure of the degree of separation between gamma and proton. Two QGSJET models show similar acceptance of background and EPOS-LHC and SIBYLL2.3c have higher values than them. This feature is consistent with the expectation from the $\pi^0$ spectra and EM fraction of those models.

\subsubsection{Effect of hadronic models on the gamma-ray sensitivity to a point-source}
 The differential sensitivity corresponds to the minimum detectable flux in each energy bin, where we set five bins in a decade. Three conditions are required for a significant detection: 1) statistical significance of the signal with respect to a background fluctuation $N_{\sigma}\geq5$, where the significance definition by Li\&Ma is used (Eq. 17 in Ref.~\cite{LiMa} with ON/OFF ratio factor $\alpha=0.2$); 2) a minimum number of signal events, $N_{\rm gamma}\geq 10$; 3) signal-to-background ratio, $N_S/N_B\geq 0.05$. These values are calculated using the number of residual events after the gamma-like event selection and angular cut. With an assumption of the observation time and gamma-ray flux, the cut position in BDT is selected in each energy bin so that all three conditions above are fulfilled with the lowest possible gamma-ray flux in the bin.
 %so that it maximizes Li\&Ma significance. Then the minimum flux which satisfies the three detection conditions is searched and defined as the differential sensitivity. 
\begin{figure}[ht]
      \centering
      \includegraphics[width=0.6\linewidth,keepaspectratio]{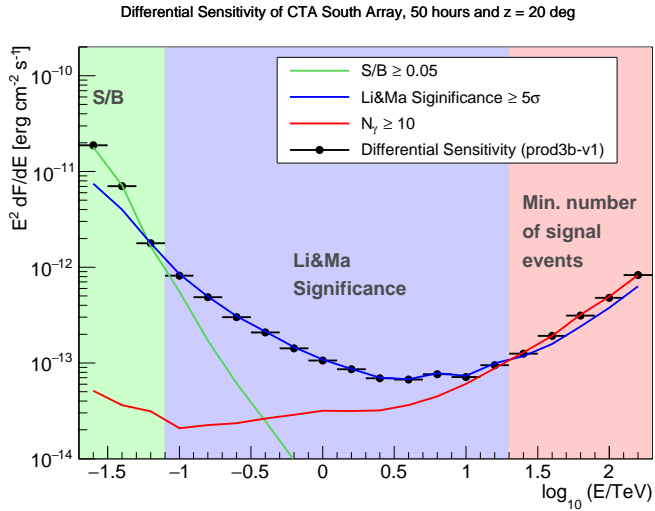}
      \caption{Differential gamma-ray sensitivity of the CTA South array to a point-source and the three conditions which determine the sensitivity. Prod3b-v1 public IRF with 50-h observation time is shown~\cite{ScienceWithCTA, CTApublicIRFprod3bv1}. QGSJET-II-03 is used in the proton simulation. Differential sensitivity depends on the energy bin size, and five bins per decade are used.}
      \label{fig:SensitivityComponents}
\end{figure}

 Figure~\ref{fig:SensitivityComponents} shows the relation of the three detection conditions with respect to the gamma-ray energy. The public IRF of the CTA South array for a point source (prod3b-v1~\cite{ScienceWithCTA, CTApublicIRFprod3bv1}) is used in the plot, with an assumption of 50-h observation time. Through the optimization of the gamma/hadron separation cut, the sensitivity curve is affected by the uncertainty in the hadronic interaction across the entire energy band, but the difference arising from this uncertainty is expected to be more clearly seen in the energy regions where the two background-related conditions determine the sensitivity: signal-to-background ratio in the $<0.1$~TeV region, and Li\&Ma significance in the 0.1~-~20~TeV region.

\begin{figure}[ht]  
    \centering
    \includegraphics[width=\linewidth,keepaspectratio]{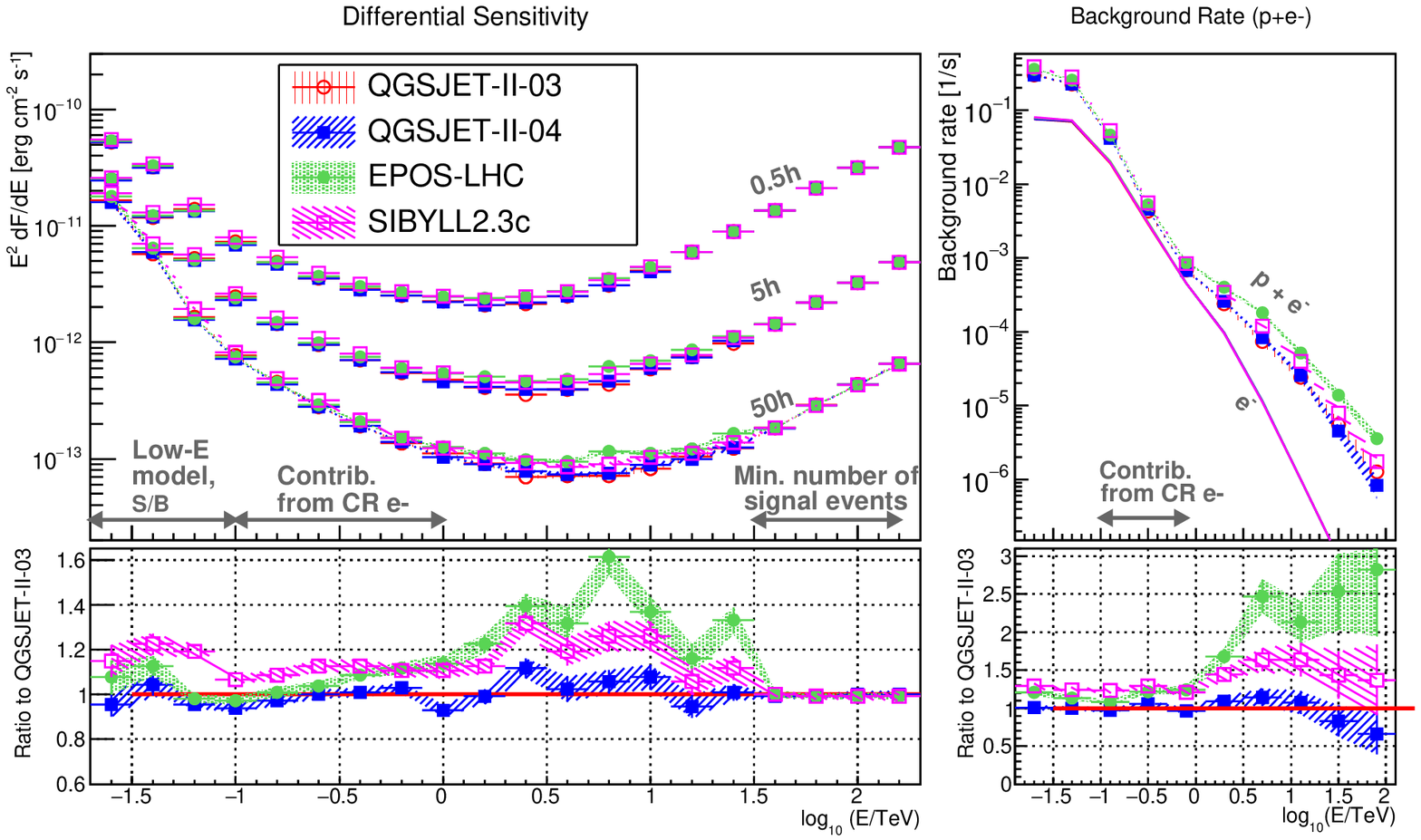}
    \caption{(Top left) Differential sensitivities for observations of a point-like gamma-ray source for three different observation times (50, 5, and 0.5 hours). Proton events produced with four interaction models are used for the background estimations. (Bottom left) Ratio of the sensitivity to QGSJET-II-03 for the 50-hour observation case. (Top right) Residual background rates, the sum of protons and electrons (p + e$^-$) and electrons only (e$^-$), for 50-hour observation time. The four electron rate curves overlap as differences between the models are small. Two energy bins are merged compared with the left figure, in order to reduce the statistical fluctuations. (Bottom right) Ratio of the residual background rates to QGSJET-II-03. The re-use of showers (20 times) in the simulation makes fluctuations look larger in the sensitivity curve (and residual background rates) compared to expectations from basic event statistics.}
    \label{fig:DiffSens}
\end{figure}

Figure \ref{fig:DiffSens} shows the gamma-ray differential sensitivity curves and the expected background rates of the CTA South array for the four interaction models. A point-like gamma-ray source is assumed, and the analysis is optimized for best point-source sensitivity while fulfilling the requirements of the angular and energy resolution.
 At the highest energies, above 30~TeV, differences among models are hardly seen, as the sensitivity curve is determined by the minimum number of signal events requirement ($N_{\gamma} \geq 10$), and hence constrained by the footprint of the telescope array. %As for the high background model (EPOS-LHC), the switching point of Li\&Ma condition and $N_{\gamma}\geq 10$ is pushed up to 30 TeV.
 
 As for the lowest energies, below 0.1~TeV, the sensitivity is determined by the $N_S/N_B\geq0.05$ condition, which might make the difference between models clearer than the Li\&Ma case (approximately proportional to $N_S/\sqrt{N_B}$ for a sufficient number of events). However, the effect of the common low-energy hadronic interaction model used where $E_{\rm true} <$ 80~GeV becomes more significant in this energy band, and as a result, the difference of the sensitivity between the models is modest.
 In the 0.1~-~30~TeV region, the Li\&Ma significance condition determines the sensitivity, but the major components of the background switch from being electron dominated (low energy side) to proton dominated around 1~TeV. Thus, differences among models are clearer in the 1~-~30~TeV region. The differences in sensitivity are caused by differences in the residual background rates which are (as a ratio to QGSJET-II-03):  $3\pm12$\% for QGSJET-II-04, $120\pm34$\% for EPOS-LHC and $54\pm10$\% for SIBYLL2.3c (here the errors correspond to the dispersion of data points in the target energy band). This causes up to $\sim$30\% differences in the gamma-ray sensitivity (as a ratio to QGSJET-II-03): $2\pm6$\% for QGSJET-II-04, $32\pm14$\% for EPOS-LHC, and $18\pm9$\% for SIBYLL2.3c. The re-use of showers (20 times) in the simulation makes fluctuations look larger in the sensitivity curve compared to expectations from basic event statistics, which is pronounced in high energy region for EPOS-LHC with its hard $\pi^0$ spectrum.

\section{Discussion}
    % GM's modifications were accepted and markups were removed by pyMergeChanges.py 
\subsection{Effect on the CTA performance of post-LHC models}
Of the four interaction models used in this work, QGSJET-II-03 is pre-LHC, and the others are post-LHC generations. QGSJET-II-04 is the same model as QGSJET-II-03 but re-parametrized to fit LHC data. Thus, the comparison between QGSJET-II-03 and 04 allows us to test the effect of LHC data on the estimation of the CTA performance. A comparison of the three post-LHC models, on the other hand,  allows us to estimate the systematic uncertainties arising from the different physics implementations in these three models.

 In the comparison of QGSJET-II-03 and 04, differences in the event rate are more significant before the selection of gamma-like events (10\%) and become smaller after the selection (3\%). As a consequence, the re-parameterization using LHC data does not affect the estimation of CTA performance as a gamma-ray detector. Lower energy collider experiments such as NA61/SHINE can provide more useful information. On the other hand, this difference between the two QGSJET-II models can affect the measurement of cosmic ray nuclei spectra by IACTs. 
 
 In the comparison of the three post-LHC models, the difference between models in event rate grows from 10\% to 50-120\% after selecting gamma-like events, the opposite trend to the QGSJET-II case. The difference in the physics implementations has a larger effect on the CTA sensitivity. The choice of the observables for the verification of the interaction model should be carefully made considering which model parameter to investigate. Further studies are needed from this point of view.

\subsection{Possibility of the verification of hadronic interaction models using CTA and current IACTs}
 CTA is expected to be able to test hadronic interaction models through various observables. By comparing  OFF-source data with MC predictions in different observables distributions, IACT arrays can contribute to the validation of the interaction models. The model verification capability correlates with the particle identification power, and CTA will have the best performance for that with its large telescope arrays.

 Validation of the hadronic interaction models using measurements from non-IACT air shower experiments has been done before (\cite{AugerHadronInt, TAmuon, KASCADEHadroinInt,TibetHadronInt}, etc.). As for IACTs, observables such as muon counts, cosmic ray rates, shower image parameters, and gamma-like event rates can be used for the validation.  A particular advantage of using the rate of gamma-like events in an IACT system is that the background events with hadronic origins are almost entirely from cosmic ray protons (see e.g., Refs.~\cite{HessElectron2008,VERITASElectron2018,SITAREK2018}).  It is consistent with the expectation that emission of very energetic $\pi^0$ close to the primary energy occurs only if the primary is a single nucleon. This feature contrasts with the model verification using muon counts, in which heavy nuclei produce more muons than protons and the cosmic ray composition is a non-negligible factor to be considered. However, even when using observables other than the gamma-like event rate, IACTs have an overlap in energy band with direct observations of cosmic ray nuclei. These independent measurements limit the effect of the uncertainty in the input cosmic ray spectra, including its heavy components.
 
\begin{figure}[t]
    \centering
    \includegraphics[width=0.7\linewidth,keepaspectratio]{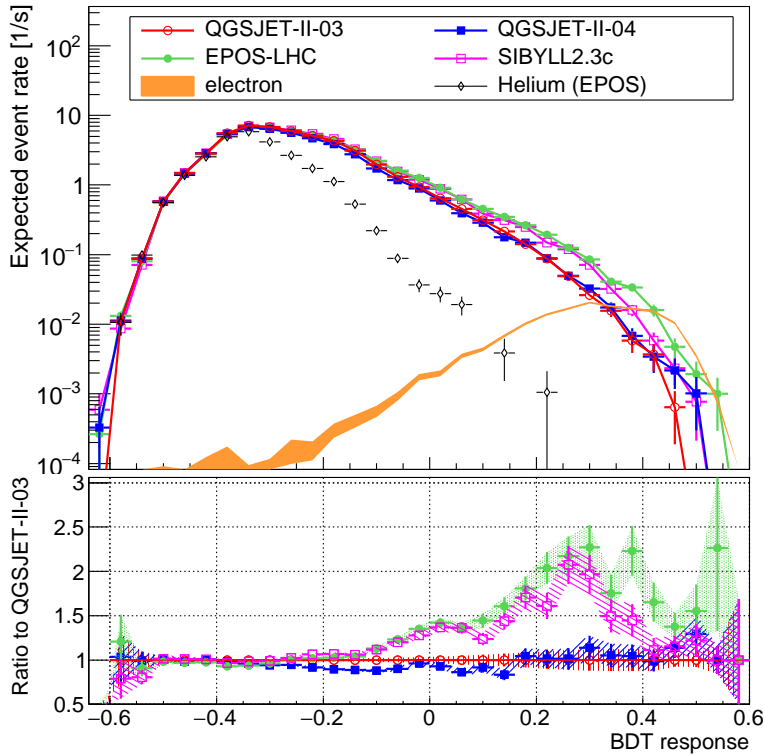}
    \caption{ (Top) BDT distributions which represent {\it gammaness} of proton events for the four interaction models and energies in the $1\leq E_{{\rm rec}_{\gamma}} \leq 10$ TeV region, where higher BDT response values correspond to more gamma-like features. Identical BDTs trained with QGSJET-II-03 are used to obtain distributions of all models for the direct comparison. The contributions from cosmic ray electrons and helium are calculated from the expected spectrum described in Sec. 1. (Bottom) Ratio to QGSJET-II-03. The expected contributions from cosmic ray electrons are taken into account, which makes the differences between models smaller for larger BDT responses. }
    \label{fig:BDTcomparison}
\end{figure}

Figure \ref{fig:BDTcomparison}~shows BDT response distributions of protons with the four interaction models in the energy range $1 <E_{{\rm rec}_{\gamma}}< 10$ TeV. Distributions of helium and electrons are also shown, where the distributions are scaled to reproduce the spectra described in Sec.2. A common BDT trained with gamma ray and QGSJET-II-03 proton events is used to directly compare the distributions (in the derivation of the sensitivity curve, BDTs are trained with each interaction model). Distributions of proton events in the gamma-like region reflect the features of the energy spectrum of $\pi^0$s near the primary proton energy and the probability of high-electromagnetic-fraction events. The two QGSJET-II models decline faster towards the gamma-like region than the other two models. The EPOS-LHC model predicts more events in the very-gamma-like region (BDT~$>0.3$) than SIBYLL2.3c. As for other particles, the helium contribution is less than 2\% in the gamma-like region (BDT~$>0$). However, we need to consider the appearance of electrons in the very-gamma-like region. Those will make the comparison more difficult at low energies due to the soft electron spectrum ($\Gamma_e=3.43$). Even taking into account the contribution from cosmic ray electrons, differences between models reach around 100\% (at BDT$=0.3$, $14\pm14$\% for QGSJET-II-04, $127\pm25$\% for EPOS-LHC, and $97\pm22$\% for SIBYLL2.3c with respect to QGSJET-II-03). This parameter is relatively a good measure for verifying interaction models; models are compared where the difference becomes large and almost free from the uncertainty in the cosmic ray nuclei composition. However, in the application to the data, this parameter may be largely affected by systematic uncertainties in the detector calibration; it uses a small fraction of events in the tail component of the broad parameter distribution. The matching between MC simulation and telescope data should be carefully examined including non-gamma-like events. Another merit of this verification method is that it needs no dedicated observation time nor special observation mode. Background events obtained in the gamma-ray observations can be reused for this purpose, as in the previous cosmic ray electron measurements by IACTs \cite{HessElectron2008,HESSelectron2009,VERITASElectron2018}. Current IACT systems are also expected to have a significant capability for model verification, though the discrimination capability is limited compared to CTA. Feedback from the data accumulated so far is encouraged. 

 The analysis method in this work follows the standard gamma-ray analysis and is not optimized for the interaction model verification. Though rates of gamma-like events turned out to be a good basis for the model verification, further development of the analysis methods specialized for the interaction model validation is possible which will increase the sensitivity to differences between the models.

\section{Conclusion}
    % GM's modifications were accepted and makrups were removed by pyMergeChanges.py

 We performed two types of Monte Carlo simulations to investigate the effect of uncertainties in hadronic interaction models on gamma-ray sensitivities for the CTA South array. Four interaction models (QGSJET-II-03, QGSJET-II-04, EPOS-LHC, SIBYLL2.3c) are considered using the air-shower simulation code CORSIKA 6.99/7.69. The first type of simulation without detector response is performed to reveal differences in secondary particle products in air showers. The second type of simulation includes the detector response and tests the difference in observables and gamma-ray sensitivities for the planned Southern site of the Cherenkov Telescope Array.
 
 As expected, the models with harder $\pi^0$ energy spectra near the primary proton energy produce more gamma-like events. Differences between models reach up to a factor 2 in the predicted residual background rate. The impact on the differential gamma-ray sensitivities is most dominant in the 1 to 30~TeV region, where the decisive condition to determine the sensitivity is the significance of the signal above the residual background fluctuations and the major background are protons. Differences of up to 30\% in the sensitivity in this energy range are observed between the hadronic interaction models. The relation between models shows an energy dependence; the highest background rate is predicted by SIBYLL2.3c below 1~TeV and switches to EPOS-LHC above 1~TeV. This relation is consistent with the results from the simulation without detector response. The two QGSJET-II models show almost similar sensitivity and residual background rates, which is also expected from their $\pi^0$ spectrum feature. Using the gamma-like proton rates in the interaction model verification has at least two advantages; the test can be done where differences between models become large, and this test is free from the uncertainty in cosmic ray nuclei composition. However, there is room to develop more sophisticated analysis methods specialized for the interaction model verification.
 These results show that IACTs are suitable detectors for the verification of the hadronic interaction models. Along with the current IACTs, CTA is the most promising detector to join the activity of verifying and improving the existing interaction models, as collider and other air shower experiments have done for many years.

%%%%%%%%%%%%%%%%%%%%%%%%%%%%%%%%%%%%%

\section*{Acknowledgments}
The digital files for the data-points of the plots in this paper are available at https://doi.org/10.5281/zenodo.4709159. Lower-level data will be available upon requests at this URL, with the approval of the collaboration.
This work was supported by JSPS Grant-in-Aid for Scientific Research Grant numbers JP17K14275 and JP20K03985. This work was partially supported by the joint research program of the Institute for Cosmic Ray Research (ICRR), the University of Tokyo.  We gratefully acknowledge financial support from the agencies and organizations listed here: 
\begin{verbatim}
http://www.cta-observatory.org/consortium_acknowledgments
\end{verbatim}
We would like to thank the computing centres that provided resources for
the generation of the Instrument Response Functions \cite{CTApublicIRFprod3bv2}. This work was conducted in the context of the CTA Analysis and Simulations Working Group. This research has made use of the CTA instrument response functions provided by the CTA Consortium and Observatory, see \cite{CTApublicIRFprod3bv1} (version prod3b-v1) for more details. This paper has gone though internal review by the CTA Consortium.
\section*{References}
\bibliography{bib_hadronicinteraction}

\end{document}